\documentclass[conference]{IEEEtran}
\IEEEoverridecommandlockouts
\usepackage{cite}
\usepackage{amsmath,amssymb,amsfonts}
\usepackage{algorithmic}
\usepackage{graphicx}
\usepackage{textcomp}
\usepackage{xcolor}

\usepackage[utf8]{inputenc}
\usepackage[T1]{fontenc}
\def\BibTeX{{\rm B\kern-.05em{\sc i\kern-.025em b}\kern-.08em
    T\kern-.1667em\lower.7ex\hbox{E}\kern-.125emX}}

\begin{document}

\title{Interactive Visualization of Proof-of-Work Consensus Protocol on Raspberry Pi\\
\thanks{
Work partially funded by grant EDU02023 (DLT Science Foundation).}
}

\author{
    \IEEEauthorblockN{
        Anton Ivashkevich\IEEEauthorrefmark{1},
        Matija Piškorec\IEEEauthorrefmark{1}\IEEEauthorrefmark{2}\IEEEauthorrefmark{3},
        Claudio J. Tessone\IEEEauthorrefmark{1}\IEEEauthorrefmark{3}
    }
    \IEEEauthorblockA{\IEEEauthorrefmark{1} Blockchain and Distributed Ledger Technologies Group, University of Zurich, Zurich, Switzerland}
    \IEEEauthorblockA{\IEEEauthorrefmark{2} Ruđer Bošković Institute, Zagreb, Croatia}
    \IEEEauthorblockA{\IEEEauthorrefmark{3} UZH Blockchain Center, University of Zurich, Zurich, Switzerland}
}

\IEEEoverridecommandlockouts
\IEEEpubid{\makebox[\columnwidth]{ \hfill} \hspace{\columnsep}\makebox[\columnwidth]{ }}

\maketitle

\IEEEpubidadjcol

\begin{abstract}
We describe a prototype of a fully capable Ethereum Proof-of-Work (PoW) blockchain network running on multiple Raspberry Pi (RPi) computers. The prototype is easy to set up and is intended to function as a completely standalone system, using a local WiFi router for connectivity. It features LCD screens for visualization of the local state of blockchain ledgers on each RPi, making it ideal for educational purposes and to demonstrate fundamental blockchain concepts to a wide audience. For example, a functioning PoW consensus is easily visible from the LCD screens, as well as consensus degradation which might arise from various factors, including peer-to-peer topology and communication latency - all parameters which can be configured from the central web-based interface.
\end{abstract}


\section{Introduction}

Beyond its original application for decentralized payment networks, blockchain technology has enabled services such as creation and ownership of digital assets, decentralized decision-making and automated financial platforms. As a result, blockchain systems have become relevant to diverse audiences across fields with and without technical background, highlighting the need for a fundamental understanding of key blockchain concepts. The consensus algorithm is one of the core blockchain components, determining the composition of the chain and managing conflicting events, ensuring that the included information is considered valid by the majority of participants. Introduced with Bitcoin, Proof-of-Work (PoW) was the first implemented consensus algorithm; it relies on probabilistic block generation, supported by \textit{miners} who simultaneously solve a cryptographic puzzle, which may lead to multiple conflicting blocks (forks) at the same blockchain height. In case of conflicts, the longest chain is eventually accepted as the canonical. 

In this paper, we introduce a demonstration setup of a fully functional Ethereum PoW blockchain network using Raspberry Pi (RPi) computers, equipped with LCD displays, and configurable parameters. The system is capable of visualizing the state of consensus within the network for educational and demonstration purposes, making it accessible to audiences without a technical background.

\begin{figure*}[htbp]
\centerline{\includegraphics[width=\textwidth]{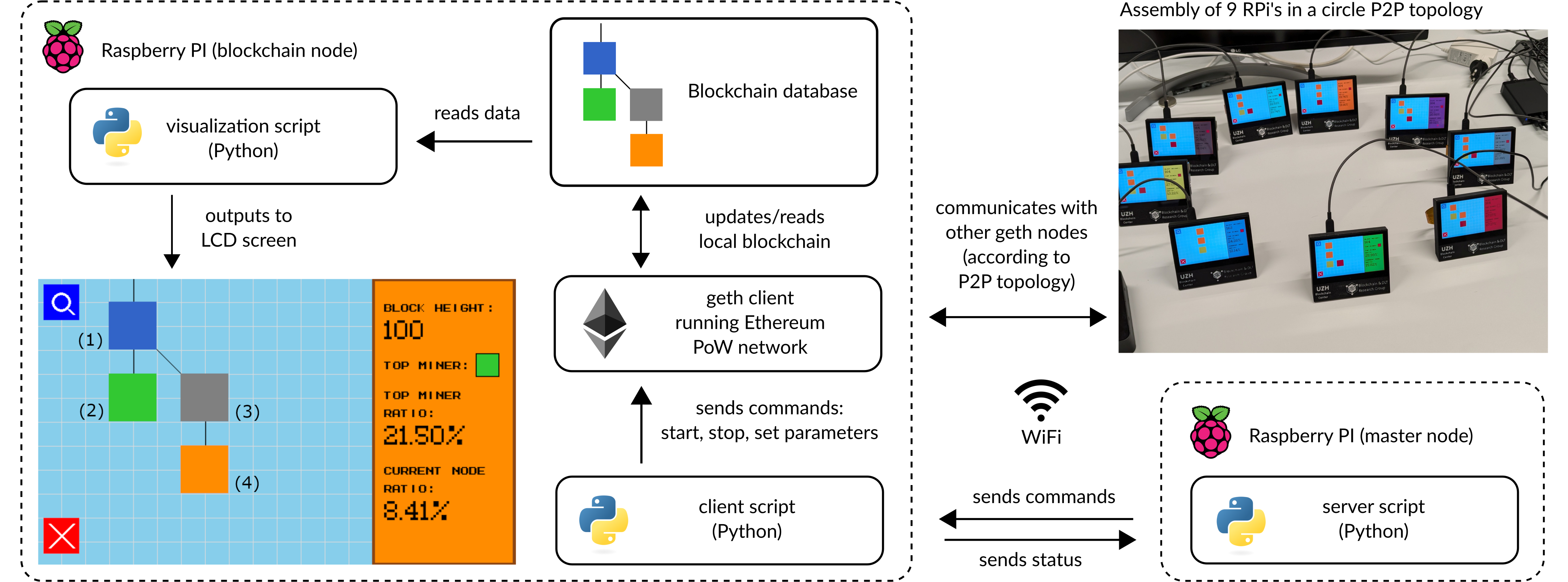}}
\caption{A schematic of the system architecture, including a screenshot of the user interface (bottom left) and a photo of the demo setup (top right), running the consensus visualization on 9 RPi SBCs. In this particular situation block propagation is still in progress, as not all nodes are displaying the latest block (dark red one). Only several nodes are in sync, while others are creating forks to the network which introduces inconsistency in the network.}
\label{fig}
\end{figure*}

\section{Description of the system}

RPi single-board computers were selected due to their affordability, wide availability, and rich connectivity options for additional devices and modules. Despite their compact size, they provide sufficient computing power to run a full PoW Ethereum node, along with the supporting network server and visualization. To further extend the compatibility of the system, our software supports all available RPi models 4 or 5 with at least 4GB RAM (8GB RAM recommended). Our demo supports screens connected via DSI, SPI and HDMI interfaces of the RPi, ensuring compatibility with most of display options. In our demo setup, we use 4.3" DSI LED screens that provide smartphone-level response times, color accuracy, and viewing angles, delivering a seamless visual experience. Deployment is done by flashing a full system image containing RaspbianOS with preconfigured dependencies on a $\mu$SD card. This simplifies system installation and configuration process and makes it possible to deploy the demo without specialized knowledge.

The demo setup comprises multiple RPi units, each running a PoW Ethereum node, along with a master RPi that manages the system through a web-based control panel. The master RPi is operated via a touch interface, allowing users to configure demo parameters and orchestrate the Ethereum nodes. The master device supplies the other units with all the required data to execute the Ethereum network, eliminating the necessity of an internet connection, therefore providing an opportunity to run the demo setup in offline environment.
The setup supports both wired Ethernet and Wi-Fi, allowing for a private network with minimal impact on communication protocols. A network configuration script automatically detects devices with predefined hostnames and dynamically establishes connections between nodes, ensuring a seamless setup.

Our project uses a modified version of the Go Ethereum client (Geth)~\cite{Geth2025}. Alongside Geth the system includes a set of Python scripts implementing communication and visualization servers, as well as supporting scripts. The source code and setup instructions are available in the GitHub repository~\cite{github}. Each RPi functions as a full node within the private Ethereum PoW network, actively participating in block production and validation.
The system allows modification of several critical parameters within the Ethereum network, including block \emph{mining difficulty} - to control the mining rate of the blocks, which naturally affects the consensus state, \emph{network latency} - the amount of time by which network communication is delayed on a node, \emph{core count} of the CPU, allocated to the miner - to adjust the hashrate of each single node, set the \emph{number} and \emph{duration} of the experiments, as well as to flexibly adjust the \emph{network topology} - the system supports fully customizable peer lists for each individual Ethereum node. The system can operate on a single node or scale to multiple nodes, simulating diverse peer-to-peer (P2P) network configurations. Therefore, the network topology can be individually adjusted without altering the physical arrangement of RPi units.

Fig.~\ref{fig} (bottom left) shows a screenshot of the user interface for a single unit running the visualization script. The right-side information panel displays the network state, including block height, the leading mining node (identified by a unique color), its block contribution percentage, and the current node's contribution. The panel’s background color uniquely identifies the node. On the left side, the visualization illustrates the blockchain state. Block, labeled (1), serves as the parent to blocks (2) and (3). A fork occurs, making block (2) an uncle block, while block (3) is included in the main chain, followed by block (4). Blocks share the same color as the corresponding node’s information panel, visually linking them to the node that mined them.

The visualization module communicates with the Ethereum node via the JSON-RPC protocol, enabling compatibility with other Ethereum Virtual Machine--based blockchains. This decoupled architecture allows the visualization to run on a separate device while connecting to the mining node over a network. Beyond visualization, each node logs detailed data from every run, preserving a complete historical record of blockchain states and operations performed at each node. This comprehensive logging provides rich data for analysis, enabling users to explore Ethereum network configurations and experiment with various blockchain scenarios.

\section{Use cases}

With its standardized and low-cost hardware, streamlined setup process, and autonomous design, the demo system can be deployed in various educational environments, catering to different levels of expertise without requiring deep technical knowledge. It effectively demonstrates blockchain consensus mechanisms, from initial consensus formation in a new network to consensus degradation over time. Adjustable network parameters enable exploration of advanced scenarios, including real P2P network simulations and specific attack types such as 51\% attacks and eclipse attacks. The use of identical hardware ensures that each mining node has a similar hashrate, enabling clear demonstrations of fair block distribution among network participants. In addition, the system’s logging functionality supports further analysis, making it a valuable tool for both educational and research purposes.

\vspace{12pt}

\end{document}